\documentclass[%
 reprint,
 amsmath,amssymb,
 aps,
  pra,
]{revtex4-2}
\usepackage{graphicx}% Include figure files
\usepackage{xcolor}
\usepackage{ragged2e}

\newcommand{\be}{\begin{equation}}
\newcommand{\ee}{\end{equation}}
\newcommand{\bee}{\begin{equation*}}
\newcommand{\eee}{\end{equation*}}

\newcommand\bra[1]{{\langle{#1}|}}
\newcommand\ket[1]{{|{#1}\rangle}}

\usepackage{mathtools}
\newcommand\braa[1]{{({#1}|}}
\newcommand\kett[1]{{|{#1})}}
\DeclarePairedDelimiterX\brakett[2]{(}{)}{#1 \delimsize\vert #2}
\newcommand{\hide}[1]{}

\DeclareMathOperator{\Tr}{Tr}

\AtBeginDocument{\renewcommand*{\d}{\mathop{\kern0pt\mathrm{d}}\!{}}}

\begin{document}

%\preprint{APS/123-QED}

\title{Effective (Floquet) Lindblad generators from spectral unwinding}% Force line breaks with \\

\author{Görkem D. Dinc}
\affiliation{%
 Technische Universität Berlin, Institut für Theoretische Physik, 10623 Berlin, Germany
}
\author{André Eckardt}%
\affiliation{%
 Technische Universität Berlin, Institut für Theoretische Physik, 10623 Berlin, Germany
}%
\author{Alexander Schnell}%
 \email{schnell@tu-berlin.de}
\affiliation{%
 Technische Universität Berlin, Institut für Theoretische Physik, 10623 Berlin, Germany
}%


\date{\today}% It is always \today, today,

\begin{abstract}
A mathematical description of the  reduced dynamics of an open quantum system can often be given in terms of a completely positive and trace preserving (CPTP) map, also known as quantum channel. In a seminal work by Wolf et al. [\textit{Phys.\ Rev.\ Lett. 101, 150402 (2008)}], it was shown that deciding whether a given quantum channel was generated from an underlying effective Markovian dynamics, with time-independent generator of Lindblad form, is generally an NP-hard problem. The difficulty is related to the fact that one has to search through all possible branches of the operator logarithm of the map, in order to identify, if any of the resulting effective generators is of Lindblad form. In this work we show that in cases where one has access to the full reduced dynamics at all previous times (the dynamical map) one can significantly facilitate the search for an effective generator by making use of Floquet theory. By performing a spectral unwinding such that the effective micromotion is minimized, the effective Floquet generator is often an excellent candidate for an effective generator of Lindblad form. This significantly reduces the complexity of the search for an effective Lindbladian in many (though not all) cases. Our results are relevant for engineering Floquet Lindbladians in complex many-body systems.
\end{abstract}

\maketitle

%\tableofcontents

%\section{\label{sec:level1}Introduction}

%\vspace{1mm}
%\newpage
%\vspace{1mm} -------------------------------------------------- \\
%\newpage

%\emph{Introduction.}---
\section{Introduction}
Digital and analog simulation of quantum many-body systems in engineered quantum systems based on photons or atoms have yielded fascinating results \cite{Altman2021,Preskill2018,Aidelsburger2011, Aidelsburger2013, Goldman2015,Eckardt2017,Clark2020,Jafferis2022, Satzinger2021,Tazhigulov2022,Leonard2023,Kwan2023,Andersen2023,Acharya2023,Bluvstein2024,Acharya2024,Xu2024,Wang2024}. Both rely in part on the advent of Floquet engineering \cite{Bukov2015,Eckardt2015,Goldman2015,Eckardt2017,Zhao2021} (which in the context of digital simulation is usually referred to as trotterization \cite{Suzuki1976,Chen2022, Wiebe2010,Sieberer2019,Childs2021,Pastori2022,Zhao2024a,Zhao2024}). The central idea is that for an isolated quantum system with time-periodic Hamiltonian $H(t)=H(t+T)$, the one-cycle time-evolution operator $U(T)$ can be rewritten as $U(T)=\exp(-iH_\mathrm{F}T)$ with a time-\emph{in}dependent Floquet Hamiltonian $H_\mathrm{F}$ with possibly novel terms that are not present in the lab Hamiltonian, e.g.~artificial gauge fields \cite{Aidelsburger2011, Aidelsburger2013,Bukov2014,Bukov2015,Goldman2015,Eckardt2017,Flaschner2016,Aidelsburger2018,Wintersperger2020} or nontrivial interactions \cite{Petiziol2021,Petiziol2024a,Sun2023,Goldman2023,Andersen2023,Petiziol2024,Poilblanc2024}.  

Recently, the question was raised \cite{Haddadfarshi2015,Schnell2020, Schnell2021} whether these ideas can be generalized to \emph{open} Floquet systems \cite{Mori2023}, i.e.~systems that interact with an environment (e.g.~by coupling the system to external photon or particle reservoirs). Interestingly, it was shown \cite{Schnell2020} that the dynamics under a time-periodic Lindbladian $\mathcal{L}(t)=\mathcal{L}(t+T)$ \cite{Schnell2020,Pastori2022,Mizuta2022,Chen2022a,Kolisnyk2024,Khandelwal2024} cannot always be understood as generated from an effective time-independent Floquet Lindbladian $\mathcal{L}_\mathrm{F}$ \cite{Haddadfarshi2015,Hartmann2017,Dai2017,Schnell2020,Schnell2021,Mizuta2021,Ikeda2021,Mizuta2022,Cemin2024}. This is due to effective non-Markovian effects that are built up during the evolution. As a result of the existence of  branches when calculating the complex logarithm of the dynamical map, deciding whether a valid Floquet Lindbladian exists is generally an NP-hard problem \cite{Cubitt2012}. This question closely relates to earlier work by Wolf et al.~\cite{Wolf_2008,Cubitt2012} where the existence of an effective Lindbladian generator was studied for a snapshot in time of a general quantum evolution $\mathcal{V}$, a completely positive and trace-preserving (CPTP) map, irrespective of where it originates from. In this work, we show that by using ideas from Floquet theory, the general problem of deciding Markovianity can be addressed more efficiently in cases where one does not only have access to such a snapshot in time, but
also to the full dynamical map $\mathcal{V}(t)$ for all times $t$.  This is relevant, e.g.~for Floquet engineering of open quantum systems, where such effective Lindbladian generators can capture the system's stroboscopic dynamics and hence be utilized in the same manner as Floquet Hamiltonians in the case of isolated systems. Our results lay the foundation for Floquet engineering of Lindbladians in extended many-body systems.

\section{Markovianity test}
%\emph{Quantum Channels, Dynamical maps.}---
The mathematical cornerstone of the theory of Markovian open quantum systems are one-parameter semigroups $\mathcal{V}(t)$ %$ \in L(L(\mathcal{H}))$, where $L(\mathcal{H})$ denotes the space of linear operators  
which are superoperators on the Hilbert space $\mathcal{H}$ with dimension $d$. The semigroup $\mathcal{V}(t)=e^{\mathcal{L}t}$, $t\ge0$,  describes an evolution of the density operator $\varrho (t)= \mathcal{V}(t)[ \varrho(0)]$ that is Markovian and consistent with a physical evolution, if the \emph{dynamical map} $\mathcal{V}(t)$ is CPTP, i.e.~a \emph{quantum channel},  at all times $t$. Lindblad~\cite{LME}, Gorini, Kossakowski and Sudarshan~\cite{Koss} have shown that a superoperator $\mathcal{L}$ is the generator of a quantum dynamical semigroup $\mathcal{V}(t)=e^{\mathcal{L}t}$, iff it can be expressed as
\be
\mathcal{L}[\cdot] = -{i}\Bigl[  H,\cdot  \Bigr]+  \sum_{i,j=1}^{d^2-1} G_{ij} \biggl( F_{i} \cdot F_{j}^\dagger -\frac{1}{2} \Bigl\{F_{j}^\dagger F_{i}   , \cdot\Bigr\}    \biggr)  , \label{eq:LF}
\ee
where we set $\hbar=k_\mathrm{B}=1$, $\{\cdot,\cdot \}$ denotes the anti-commutator, $H=H^\dagger$ the Hamiltonian, $G_{ij}$ a positive semidefinite Kossakowski matrix, and $F_{i}$ a basis of traceless operators. $\mathcal{L}$ is called a \emph{Lindblad  superoperator} or \emph{Lindbladian}. 

%Embedding problem (Floquet schon hier)

%\emph{Markovianity test.}---
Let us turn to the converse question and ask ``Given an arbitrary quantum channel $\mathcal{V}$ (say at time $T$) is there an effective generator 
\be
\mathcal{S}=\log(\mathcal{V}) / T
\ee
that has Lindblad form as in Eq.~(\ref{eq:LF})?''. Due to the multi-valuedness of the complex logarithm, this question turns out to result in an NP-hard problem in general, also called the \textit{Markovianity problem}~\cite{Wolf_2008,Cubitt2012}.~A criterion for Markovianity in this sense was proposed by Wolf et al.~\cite{Wolf_2008,Cubitt2012}: A generator $\mathcal{S}$ has Lindblad form, iff (i) for all Hermitian $\kappa=\kappa^\dagger$, $\mathcal{S}$ preserves Hermiticity, i.e.~$\mathcal{S}[\kappa] = \bigl(\mathcal{S}[\kappa] \bigr)^\dagger$, and (ii) $\mathcal{S}$ is \textit{conditionally completely positive} (CCP), i.e. 
\be
 (\mathbf{1}_{d^2} - \Sigma )  \mathcal{S}^\Gamma (\mathbf{1}_{d^2} - \Sigma )  \ge 0   .  \label{eq:ccp}
\ee
Here, $\Sigma = \ket{\Omega}\bra{\Omega}$ is the projector onto the maximally entangled state $\ket{\Omega} =  \frac{1}{\sqrt{d}}  \sum_{j=1}^{d}  \ket{j} \otimes \ket{j}$ of the system and an ancilla of same dimension $d$ and
\be
    \mathcal{S}^\Gamma  = d \cdot (\mathcal{S} \hspace{0.5mm} \otimes \hspace{0.5mm} \mathbf{1}_{d}) [ \Sigma ]
\ee
denotes the Choi representation of $\mathcal{S}$. Since the complex logarithm is not uniquely defined,
we need to check if there is a branch of the operator logarithm of $\mathcal{V}$ that satisfies condition (i) and (ii). 

To do this,
we introduce the operator scalar product $\brakett{A}{B}=\Tr[A^\dagger B]$.
%on $L(\mathcal{H})$. 
By also defining an orthonormal operator basis $\{ F_\alpha  \}_{\alpha=1}^{d^2}$, 
%on this Hilbert space $L(\mathcal{H})$, 
we can represent quantum channels $\mathcal{V}$ as matrices $\hat{\mathcal{V}}$, with their corresponding matrix elements $\hat{\mathcal{V}}_{\alpha \beta}=\braa{F_\alpha}\mathcal{V}\kett{F_\beta}$, which allows for the identification $\varrho=\ket{i}\bra{j} \rightarrow \kett{\varrho} =\kett{i,j}$, known as vectorization of (density) matrices. A useful operation is $\mathbb{F} \bigl[ \sum_{i,j} c_{ij} \kett{i,j} \bigr] = \sum_{i,j} c_{ij}^* \kett{j,i}$ \cite{Wolf_2008}. %, where $\mathbb{F}$ is called the flip operator and relates the spectral projectors of complex conjugated eigenvalue pairs. 
Assuming the map $\hat{\mathcal{V}}$ to be non-defective, which is generally justified for Markovian maps and quantum channels \cite{Cubitt2012}, we first cast it into Jordan normal form \cite{Wolf_2008}, 
\be
    \hat{\mathcal{V}}=  \sum_{r} \lambda_r\kett{r_r}\braa{l_r} + \sum_{c=1}^{N_c} \bigl[\lambda_c \kett{r_c}\braa{l_c} + \lambda_c^* \ \mathbb{F} \kett{r_c}\braa{l_c}\mathbb{F} \bigr],
\ee   
     with $r$ and $c$ indexing the real and complex eigenvalues, respectively, and ${N_c}$ the total number of complex conjugated pairs. The logarithm yields a family of possible generators of $\mathcal{V}$ \cite{Wolf_2008,Schnell2020}, 
\be
\hat{\mathcal{S}}_{\vec{x}} = \hat{\mathcal{S}}_0 +  i\frac{2\pi}{T}  \sum_{c=1}^{N_c} x_c  \Bigl[ \kett{r_c}\braa{l_c}-  \mathbb{F} \kett{r_c}\braa{l_c}\mathbb{F} \Bigr] \label{eq:LG}, 
\ee
with $\hat{\mathcal{S}}_0$ stemming from the principal branch of the logarithm and a vector $\vec{x}=\{ x_1 ,   x_2,... , x_{N_c}  \} \in  \mathbb{Z}^{N_c}$  of integers labeling the possible branches of the complex logarithm.
%If all real eigenvalues $\lambda_r$ of $\hat{\mathcal{V}}$ are positive \textcolor{orange}{and non-degenerate}, % and, a
%all of these generators $\mathcal{S}_{\{\vec{x}\}}$ preserve hermiticity by construction and hence fulfill condition (i).
Note that, to arrive at Eq.~\eqref{eq:LG}, we implicitly assume that condition (i) is fulfilled \cite{Wolf_2008,Schnell2020}, i.e.~that  all real eigenvalues $\lambda_r$ of $\hat{\mathcal{V}}$ are positive \footnote{To fulfil condition (i), negative real eigenvalues $\lambda_r<0$ can only occur with an even degeneracy, such that one can reinterpret them as complex conjugated pairs}.
Naively, to test condition (ii), we have to inspect all branches, i.e., a countably infinite number of combinations of $N_c$ integers. Nevertheless, by use of methods from integer programming, the problem can be reduced to a smaller set of possible integers \cite{Cubitt2012}, or alternatively one can apply machine-learning algorithms \cite{Volokitin2022}. 

However, it was shown \cite{Cubitt2012} that deciding Markovianity is generally an NP-hard problem. This is especially detrimental in the case of interacting many-body systems where already the underlying Hilbert space $\mathcal{H}$ grows exponentially with system size $L$. 
Consider for example a spin chain of length $L$, with $d=\mathrm{dim}(\mathcal{H})=2^L$. %so the maximum number of complex pairs (since $\lambda_r=1$ always exists \cite{Chen2024}) is given by $N_c \leq 2^{2L-1}-1$. 
If we would just naively check the two closest branches around the principle branch, i.e.~$x_c \in \lbrace -1, 0, 1\rbrace$, we already find super exponential scaling $3^{2^{2L-1}-1}$ of the number of branches for which condition (ii) has to be checked. 

In the case that no branch gives rise to a valid Lindbladian, one can measure the distance to Markovianity  and find the branch that is closest to a Markovian evolution. We use a measure proposed by Wolf et al.~\cite{Wolf_2008} that is based on adding a noise term $\chi \mathcal{Z}$ of strength $\chi$ to the generator  $\mathcal{S}_{\{\vec{x}\}}$, where $\mathcal{Z}$ is the generator of the depolarizing channel.
%$\exp({T \chi \mathcal{Z}}) [\varrho]=e^{-\chi T}\varrho + (1-e^{-\chi T})\mathbf{1}/d$.
The distance from Markovianity is defined as~\cite{Wolf_2008} 
\be
\mu = \min_{\{\vec{x}\} \in  \mathbb{Z}^{N_c}} \min \ \Bigl\{ \chi \ge 0  \big| \mathcal{S}_{\{\vec{x}\}}+\chi \mathcal{Z} \text{ is CCP}  \Bigr\},   \label{eq:measure}
\ee
i.e. the minimal strength needed, such that the generator $\mathcal{S}_{\{\vec{x}\}}+\chi \mathcal{Z}$ is Lindbladian. %, i.e. fulfills condition (i) and (ii). %The measure therefore reads %. More precisely, $\mu$ is defined as \cite{Wolf_2008}
Details on the definition and how to calculate this measure $\mu$ are laid out in Refs.~\cite{Wolf_2008,Schnell2020}.

\section{Systems with access to full time-dependent map}
%\emph{Systems with access to full time-dependant map.}---
We describe  a strategy that allows to address (and often solve) this problem in cases, where additionally to $\mathcal{V}$ at time $T$, the full dynamical map $\mathcal{V}(t)$ is known also for times $t \in [0,T]$. To this end, we draw a connection to Floquet theory by continuing $\mathcal{V}(t)$ periodically for $t>T$, e.g.~$\mathcal{V}(t)=\mathcal{V}(t-T)\mathcal{V}(T)$ for $T<t\leq 2T$ and so on (a similar strategy was employed in the context of isolated systems for the purpose of adiabatically preparing eigenstates of time evolution operators~\cite{Unal2019}). Then, the time evolution can effectively be understood as generated from a time-periodic generator $\mathcal{G}(t)=\mathcal{G}(t+T)$ (not necessarily of Lindblad form) since
\be
    \partial_t \varrho(t) = [\partial_t\mathcal{V}(t)] \mathcal{V}(t)^{-1} \varrho(t) =\mathcal{G}(t) \varrho(t),
    \label{eq:eff-gen}
\ee
where we assume that $\mathcal{V}(t)$ is differentiable and invertible.

Let us briefly discuss Floquet theory for isolated (and open) systems. For systems with time-periodic Hamiltonian $H(t)=H(t+T)$ (generator $\mathcal{G}(t)$),
the fundamental solutions can be written in terms of Floquet states $\ket{\psi_\alpha(t)}=\exp(-i\varepsilon_\alpha t) \ket{u_\alpha(t)}$  $\bigl(
\kett{\varrho_\mu (t)} = \exp(-i\Omega_\mu t) \kett{\Phi_\mu (t)}\bigr)$ with time-periodic Floquet modes $\ket{u_\alpha(t)}=\ket{u_\alpha(t+T)} \ \bigl(\kett{\Phi_\mu (t)}=\kett{\Phi_\mu (t+T)} \bigr)$  \cite{Schnell2021,Chen2024,Eckardt2015} and quasienergies $\varepsilon_\alpha$ \cite{Eckardt2015} (eigenvalues $\Omega_\mu$). In case of a purely coherent evolution, $\mathcal{G}(t)=-i[H(t), \cdot]$, the $\Omega_\mu$ correspond to \emph{quasienergy differences} (a term that we will continue to use also for the dissipative case). %Here, we still refer to the  $\Omega_\mu$ as quasienergy differences, which are complex in general.
Note that the quasienergies $\varepsilon_\alpha$ (quasienergy differences $\Omega_\mu$) are not uniquely defined and can be redefined under the gauge transformation $\varepsilon_\alpha \rightarrow \varepsilon_\alpha +m\omega$ and $\ket{u_\alpha(t)} \rightarrow e^{im\omega t}\ket{u_\alpha(t)} \ \bigl(\Omega_\mu \rightarrow \Omega_\mu +m\omega$ and $\kett{\Phi_\mu (t)}  \rightarrow 
e^{im\omega t}\kett{\Phi_\mu (t)} \bigr)$, where $\omega=2\pi/T$ and $m \in \mathbb{Z}$. In the isolated case, picking a specific gauge is equivalent to fixing the branch of the complex logarithm when solving for $H_\mathrm{F}=i\log[U(T)]/T$. Stroboscopically, at $t=0, T, 2T, \dots$ the time evolution operator $U(T)=\exp(-iH_\mathrm{F}T)$ is described by the effective time-independent Floquet Hamiltonian 
\be
    H_\mathrm{F}=\sum_\alpha \varepsilon_\alpha\ket{u_\alpha(0)}\bra{u_\alpha(0)}
    .
\ee
In the dissipative case,
for a given time $t$, we can decompose the dynamical map $\hat{\mathcal{V}}(T)$ as 
\be
    \hat{\mathcal{V}}(T)=\sum_\mu e^{-i\Omega_\mu T} \kett{\Phi_\mu (T)} \braa{\Tilde{\Phi}_\mu (0)},
\ee
with the left Floquet modes $\braa{\Tilde{\Phi}_\mu (t)}$ being different from the right ones for dissipative systems with a non-hermitian generator \cite{Schnell2021}. 
% By construction, 
%$
%\kett{\Phi_\mu (t)} = e^{i\Omega_\mu t} \hat{\mathcal{V}}(t)\kett{\Phi_\mu (0)}
%$
%and
%$
%\braa{\Tilde{\Phi}_\mu (t)} = \braa{\Tilde{\Phi}_\mu (0)} \hat{\mathcal{V}}(t) e^{i\Omega_\mu t}, 
%$
%which allows for numerical computation of the Floquet modes $\kett{\Phi_\mu (t)}$ after diagonalization of the map $\mathcal{V}(T)$. 
In this way, we can immediately identify the effective generator as~\cite{Schnell2021}
\be
\hat{\mathcal{S}}=-\sum_\mu i\Omega_\mu  \kett{\Phi_\mu(0)} \braa{\Tilde{\Phi}_\mu(0)},
\ee
with the freedom of choosing the Floquet gauge of the quasienergy differences $\Omega_\mu$ indicating the 
freedom of choosing a branch of the logarithm in Eq.~\eqref{eq:LG}. %Henceforth, we  write $\kett{\Phi_\mu (0)} \equiv \kett{\Phi_\mu } $ and $\braa{\Tilde{\Phi}_\mu (0)}\equiv\braa{\Tilde{\Phi}_\mu } $.
%since the Floquet modes are periodic in time.
%are the signs in the decompositions (especially in the exponents) all right?
\begin{figure}
\includegraphics[width=0.9\linewidth]{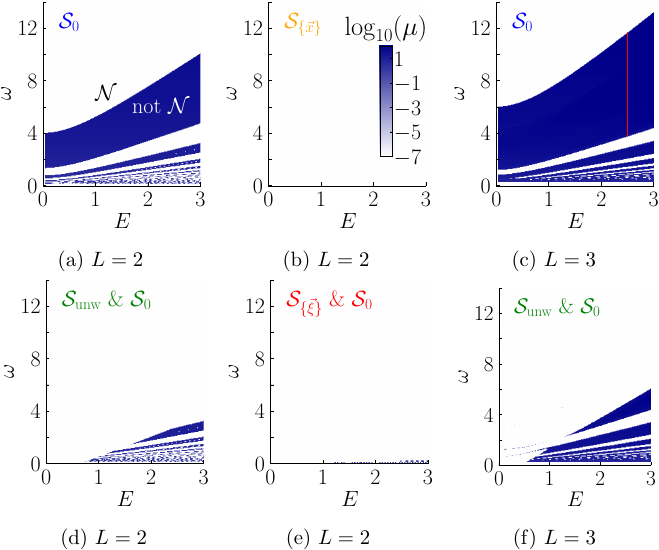}
\caption{%\justifying
(a)-(f) Distance from Markovianity $\mu$ (which here equals to distance from von Neumann form $\mathcal{N}$) of the effective generator
of the one-cycle evolution superoperator as a function of driving
strength $E$ and frequency $\omega$, for no dissipation, $\gamma=0$, coupling strength $w=0.1$, and chain lengths of (a),(b),(d),(e) $L=2$ and (c),(f) $L=3$. 
To avoid numerical artifacts stemming from degeneracies in the spectrum of $\mathcal{V}(T)$, we set $\Delta_1=1.01,\Delta_2=1.00,\Delta_3=0.98$.
%To lift degeneracies in the spectrum \textbf{\textcolor{orange}{to avoid numerical artifacts}}, we set  $\Delta_1=1.01,\Delta_2=1.00,\Delta_3=0.98$.
}
\label{fig:2spins}
\end{figure}

Let us now, for illustrative purposes, regard the evolution with a time-independent generator, $\partial_t \varrho(t)=\mathcal{L} \varrho(t)$, such that $\mathcal{V}(T)=\exp({\mathcal{L} T})$, as an effective time-periodic problem with arbitrary period $T$. If we choose a large $T$ such that $\omega=2\pi/T$ is smaller than the imaginary part of some of the eigenvalues
of $\mathcal{L}$, we observe that the generator $\mathcal{S}_0$ that we obtain from the principle branch of $\log(\mathcal{V}(T))/T$ is \emph{not} identical to $\mathcal{L}$ but there occurs a ``winding'' of the eigenvalues of the Floquet generator $\mathcal{S}$ ``around'' the first  \textit{Floquet Brillouin zone}, $\mathrm{Re}[ \Omega_\mu] \in [-\omega/2, \omega/2)$, since $\exp(-i \Omega_\mu T)$ has the shape of a circle.
This winding turns out to make the previously outlined procedure of searching around the principle branch inefficient, even for a fully coherent time evolution with \emph{no} dissipation: To demonstrate this, we consider a circularly driven spin chain of length $L$
\begin{align}
    & H(t) = \sum_{\ell=1}^L \frac{\Delta_\ell}{2}\sigma_z^\ell + \sum_{\ell=1}^{L-1}w\sigma_x^\ell \sigma_x^{\ell +1} + H_{\text{drive}}(t) \\ \label{eq:spinchain}
     & H_{\text{drive}}(t) =  \sum_{\ell=1}^L E \bigl[ \cos(\omega t) \sigma_x^\ell - \sin(\omega t) \sigma_y^\ell \bigr] ,
\end{align}
with level splitting $\Delta_\ell$, coupling strength $w$ between neighbouring spins, driving strength $E$ and -frequency  $\omega$.
As we show in Fig.~\ref{fig:2spins}(a) and (c), even if we calculate the dynamical map $\mathcal{V}(T)$ for an evolution with time-dependant von-Neumann generator $\mathcal{G}(t)=-i[H(t), \cdot]$, the generator $\mathcal S_0$ of the principle branch is \emph{not} necessarily of von-Neumann form $\mathcal{N}$ (and neither of Lindblad form), see blue areas. This is in stark contrast to our knowledge of the existence of the Floquet Hamiltonian $H_\mathrm{F}$, i.e.~$\mathcal S=-i[H_\mathrm{F}, \cdot ]$ is a valid effective Lindblad generator. Only after also searching around the first two neighbouring branches, i.e.~$x_c \in \lbrace -1, 0, 1\rbrace$, one finds again a valid von Neumann generator, as we show in Fig.~\ref{fig:2spins}(b). 
However, for every point in the phase diagram, a maximal number of $3^{2^{2L-1}-1}$ branches has to be checked for condition (ii). %In the case of a spin chain of length $L=2$ this can still be computed in a reasonable time. 
Due to the super exponential scaling, for $L=3$ ($L=4$) we would already have a maximal number of $\approx 6 \cdot 10^{14}$ ($ \approx  3 \cdot   10^{60}$) branches to consider, simply to avoid the winding problem (cf.~yellow line in Fig.~\ref{fig:L3L4}(a)). 
In order to avoid this problem, we propose a different approach which is based on a strategy for unwinding the quasienergy spectrum.

%\emph{Spectral unwinding.}---
\section{Spectral unwinding}
For undriven systems it is straightforward to define an ``unwinding'' procedure that undoes this winding into the first  Floquet Brioullin zone: 
%Why this is the case is demonstrated in Appendix A. To undo the folding or to \textit{unfold} our generator,  we need the lost information about the unitary rotations.~This number of unitary rotations on the unit circle is contained in the \textit{micromotion}, i.e.~the time dependency of  $\kett{\Phi_c(t)}$ in Eq.~(\ref{eq:td}). The aim is to make the micromotion as little as possible, such that the Floquet modes become nearly constant.
Since the Floquet modes $\kett{\Phi_c(t)}$ with corresponding complex quasienergy differences $\Omega_c$ are time-periodic, we can expand them in a Fourier series  
%\be
\be
    \kett{\Phi_c (t)} = \sum_{n \in  \mathbb{Z}} e^{i \omega nt} \kett{\Phi_c^{(n)}}  ,
\ee
%\ee
and compute the trace norm for all of their Fourier components $|| \kett{\Phi_c^{(n)}} || = \Tr\sqrt{\Phi_c^{\dagger(n)} \Phi_c^{(n)}}$.
%The number of rotations that the phase performs on the unit circle during period T is then given by the frequency $n$ whose Amplitude in the Fourier spectrum is maximally peaked. More accurately, 
For every mode $c$ we determine  
\be
x_c^{(\text{max})}= {\arg\max}_{x\in \mathbb{Z}} \{ || \kett{\Phi_c^{(x)}} || \} .
\ee
Then, we shift the  $\Omega_c$ by this number of quanta $x_c^{\text{(max)}}$ giving the `unwound' Floquet generator 
\begin{equation}
\begin{split}
\hat{\mathcal{S}}_\text{unw}
= \hat{\mathcal{S}}_0 
+ i \omega \sum_{c=1}^{N_c}  x_c^{(\text{max})} \Bigl[ 
\kett{\Phi_c(0) } \braa{\Tilde{\Phi}_c (0)} \\
- \mathbb{F} \kett{\Phi_c(0)} \braa{\Tilde{\Phi}_c (0)} \mathbb{F} 
\Bigr]
\end{split}
\end{equation}
%where $\hat{\mathcal{S}}_0$ again denotes the generator of the principle branch.
In the undriven system, this gauge transformation removes the micromotion and transforms the Floquet modes into the static eigenmodes of the time-independent generator $\mathcal{G}$ \cite{Eckardt2015,Leskes2010}.
We now show that this procedure also significantly improves the search for an effective generator in the case of a time-dependent generator $\mathcal{G}(t)$.  %and in this regard, we go back to our spin chain model. 
In Fig.~\ref{fig:2spins}(d) and (f) we show the phase diagrams we obtain from checking   the principle branch $\mathcal S_0$ and the unwound generator $\mathcal{S}_{\text{unw}}$ for chain lengths $L=2$ and $L=3$. Here, between $\mathcal S_0$ and $\mathcal{S}_{\text{unw}}$, we choose the generator with shortest distance from Markovianity $\mu$. In %the Supplemental Material (SM) \cite{SM}
Appendix \ref{sec:app1}, we plot the phase diagrams of $\mathcal{S}_{0}$ and $\mathcal{S}_{\text{unw}}$ separately. By using the procedure above, in comparison to Fig.~\ref{fig:2spins}(a) and (c), we find that specifically in the high-frequency regime we obtain a lot more points $(E, \omega)$ yielding generators of von Neumann form. 
%In Fig.~\ref{fig:2spins} we plot the distance from Markovianity $\mu$. 
(In App.~\ref{sec:app-vN}, %SM \cite{SM}
 we introduce a measure to determine the distance from the von-Neumann form and show that both measures almost coincide for our problem.) \\
\indent However, as we observe in Fig.~\ref{fig:2spins}(d),(f) for low  frequencies, this method of unwinding fails. This is expected, since in this regime the peaks in the Fourier spectra can become more broadly distributed over several Fourier modes and indistinctly peaked, which we illustrate in %the SM \cite{SM} 
App.~\ref{sec:app-Four},
by specifically analyzing the Fourier spectra of $|| \kett{\Phi_c^{(n)}} ||$ at points, where $\mathcal{S}_{\text{unw}}$ fails and succeeds in giving a valid generator of von Neumann form. To overcome this issue, one has to include additional frequency peaks for the search in this regime. 
In the following, we outline how to find well-chosen  candidates for effective generators to  test for Markovianity, other than $\mathcal S_0$ and $\mathcal{S}_{\text{unw}}$. To this end, we set $x_c^{(\text{max,0})}=x_c^{(\text{max})}$ and choose a number $\eta \in [0,1]$, which determines at which amplitude ratio we want to include the neighbouring peaks $x_c^{(\mathrm{max},i>0)}$ into the possible branch combinations. %and the main peak $x_c^{(\text{max,0})}$ them %the neighbouring peaks 
%into the generation of branch combinations. 
The resulting family of generators $\hat{\mathcal{S}}_{\{\vec{\xi}\}}$ we consider is given by
%\be
%    \hat{\mathcal{S}}_{\{\vec{\xi}\}}
%    =  \hat{\mathcal{S}}_0 + i \frac{2\pi}{T} \sum_{c=1}^{N_c}  \xi_c \cdot  \bigl( \kett{\Phi_c } \braa{\Tilde{\Phi}_c }  -\mathbb{F} \kett{\Phi_c} \braa{\Tilde{\Phi}_c } \mathbb{F} \bigr)   ,  \label{eq:modMar}
%\ee 
%with the branches according to Eq.~(\ref{eq:LG}) but here,
%the integers $ \xi_c $ are not any arbitrary integers, but the frequencies of the main and the included neighbouring peaks in the Fourier spectra of every Floquet mode 
%\be
\begin{align}
%\begin{split}
\{\vec{\xi}\} = \Bigl\{  ( \xi_i ) \in  \mathbb{Z}^{N_c}   \ \big| \
& \xi_{c}  \in \{x_{c}^{(\text{max,0})},...,x_{c}^{(\text{max,$z_{c}$})}\}  \Bigr\},
%\end{split}
\label{eq:branches-unf}
\end{align}
%\ee
where for every pair of complex modes $c$,  we include the $z_{c}$  neighbouring Fourier peaks %(at $x^{(\text{max,i})}$) 
with amplitudes %that are on the order of the amplitude of the main peak according to the criterion
%(at $x^{(\text{max,0})}$)
$%\be
 {A^{(\text{max,i})}_{c}}/{A^{(\text{max,0})}_{c}} \ge \eta    ,
$ %\ee
with $A^{(\text{max},i)}_c = || \kett{\Phi_c^{(x^{(\text{max,i})})}} ||$.
%=\Tr\sqrt{\Phi_c^{\dagger(x^{(\text{max,i})})} \Phi_c^{(x^{(\text{max,i})})}}$. 
We additionally introduce  a cutoff $N_b$ for the maximal number of branches considered, i.e.~$z_c\leq N_b$. %and only consider the peaks $i=1,...,N_b$. 
This gives rise to a systematic search around the unwound generator $\mathcal{S}_{\text{unw}}$, instead of searching around the generator of the principle branch $\mathcal S_0$, which can drastically reduce the required numerical effort.
%, which essentially results in a smaller set of Floquet generators $\mathcal{S}_{\{\vec{\xi}\}}$ for which we test for Markovianity. 
In Fig.~\ref{fig:2spins}(e) we apply this modified Markovianity test for a spin chain of length $L=2$ and parameters $N_b=2, \eta=0.7$. We observe that we almost fully recover valid von Neumann generators. Note that for $L\ge 3$ this choice of $N_b$ and $\eta$ will not suffice to recover valid von Neumann generators in every point of the phase diagram. Here, to minimise the numerical effort, for a given point, one can search for the smallest $N_b$ and largest $\eta$ which yield a generator of von Neumann form. In %the SM \cite{SM}
App.~\ref{sec:app-comp}, for $L=3$ we have set $N_b=1$ and varied $\eta$ to find an optimum along the vertical red line at $E=2.5$ in Fig.~\ref{fig:2spins}(e). %For details on the specific values of $\eta$ for different points on the line we refer to the Supplemental Material.

\begin{figure}
\includegraphics[width=0.9\linewidth]{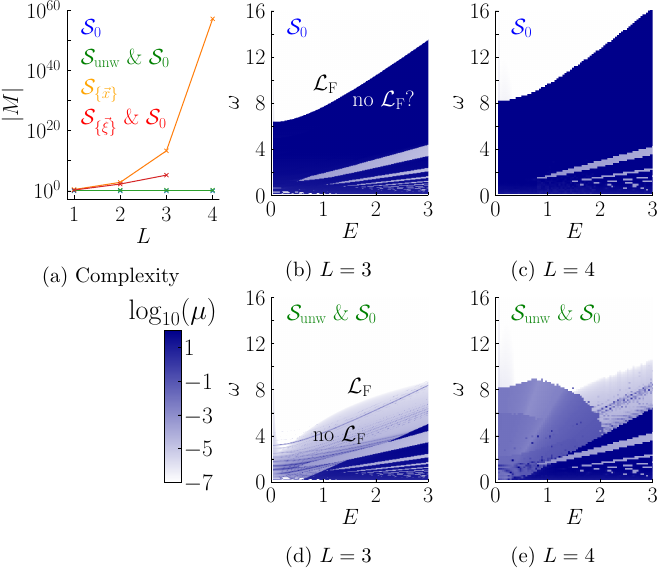} 
    \caption{%\justifying
    (a) Cardinality $|M|$ of the set $M$ of possible branch combinations for the different methods and lengths $L$. (b)-(e) Distance from Markovianity $\mu$ of the effective generator
of the one-cycle evolution superoperator as a function of driving
strength $E$ and frequency $\omega$, for weak dissipation $\gamma = 0.01$, bath temperature $\mathcal{T}=1$ and chain lengths of (b),(d) $L=3$  and (c),(e) $L=4$. Other parameters $\Delta_1=1.4,\Delta_2=1.1,\Delta_3=0.7, \Delta_4=0.9, w=0.01$.}
    \label{fig:L3L4}
\end{figure}

To compare the reduction in complexity of our method to the naive procedure of searching around the principle branch, in Fig.~\ref{fig:L3L4}(a) we plot the cardinality $|M|$ of  the set $M$ of all possible branch combinations for which we test for Markovianity. %Here, for each method, we choose the smallest possible $|M|$ for which we still find the right branch at any given point $(E,\omega)$. 
%For the standard Markovianity test, the size of $M$ is dependent on the included neighboring branches $x_c$. For the modified Markovianity test the size of $M$ depends on our choice of $N_b$ and $\eta$, which we choose such that we minimise the included branches but still get a proper von Neumann generator. 
For the principle branch $\mathcal{S}_0$ we always have $|M|=1$ (blue line). The generators of the principle branch yield poor results due to the winding of quasienergy differences. Including the first two neighbouring, i.e.~$x_c \in \lbrace -1, 0, 1\rbrace$, fixes this problem, but greatly increases the complexity as we see for the orange line. 
On the other hand, trying to revert the winding by only including $\mathcal{S}_0$ and $\mathcal{S}_{\text{unw}}$, we have $|M|=2$  (green line). 
Using the modified Markovianity test of Eq.~\eqref{eq:branches-unf}, the red line in Fig.~\ref{fig:L3L4}(a) shows the minimal cardinality needed to obtain a fully von Neumann phase diagram for $L=2$, as well as to make the points on the red line in Fig.~\ref{fig:2spins}(c) fully von Neumann for $L=3$. 
For  $L=3$ we have $|M| \approx 10^5$ on average (see %SM \cite{SM} 
App.~\ref{sec:app-comp} for details). % to make the red line fully von Neumann. 
Searching around the principle branch requires $|M| \approx 10^{13}$. This shows that searching around the unwound generator dramatically  decreases the complexity (as long as we are not at low frequencies, where the
peaks in the Fourier spectra may spread across multiple Fourier modes). For systems with large Hilbert space dimensions, where it is numerically
impossible to test all branch combinations when considering e.g.~the principal and first two neighboring branches, i.e.~$x_c \in \lbrace -1, 0, 1\rbrace$, the conventional method would require to blindly guess a branch combination with no physical or mathematical justification, while our method provides a systematic way of
searching for new generators in cases where the principle branch $\mathcal{S}_0$ and unwound generator $\mathcal{S}_{\text{unw}}$ yield poor results. Nevertheless, even for the modified method, the complexity still grows fast with the dimensionality of the problem, which is why for bigger systems one might only be able to include the principle branch $\mathcal{S}_0$ and the unwound generator $\mathcal{S}_{\text{unw}}$ (or alternatively, combine our method with the existing machine learning method of Ref.~\cite{Volokitin2022}).

\section{Dissipative case}
%\emph{Dissipative case.}---
In the case of a coherent time evolution we can always compute a Floquet Hamiltonian $H_\mathrm{F}$ and then obtain the effective time-independent Floquet generator $\mathcal S=-i[H_\mathrm{F}, \cdot ]$, which always has von Neumann form. 
Such a bypass, however, is not possible in the case of dissipative systems where we search for effective time-independent generators of Lindblad form, i.e.~Floquet Lindbladians \cite{Schnell2020,Schnell2021}. 
Here we have to compute $\log(\mathcal{V}(T))/T$ and test these generators for Markovianity \cite{Schnell2020}. 
If we do not change the Hamiltonian but add very weak dissipation, we can see exactly the same non-Markovian areas arising in the principle branch,
%(due to the winding of the eigenvalues of $\log[\mathcal{V}(T)]/T$), 
cf.~Fig.~\ref{fig:L3L4}(b).   
Here, we add dissipation described by Lindblad jump operators for the quantum-optical master equation 
\be
   L^\ell_{kq} = \sqrt{R^\ell_{kq}} \ket{\psi_k}\bra{\psi_q},
\ee
where, for simplicity, we neglect the impact of the driving term on the dissipator. Hence, $\ket{\psi_k}, E_k$ are the many-body eigenstates and -energies of the undriven Hamiltonian for $E=0$, respectively.
The corresponding jump rate for contact with a reservoir at temperature $\mathcal{T}$ at site $l$ reads
\be
    R^\ell_{kq} = 2\pi \gamma^2 \vert \bra{\psi_k} \sigma_x^\ell \ket{\psi_q} \vert^2 g(E_k -E_q)
\ee
with $
    g(E) = E/(e^{E/\mathcal{T}}-1)
$, assuming an ohmic bath~\cite{breuer2002}.

By summing over all jump operators we obtain the total time-periodic Lindbladian 
\be
    \mathcal{L}(t)[\cdot] = -{i}\bigl[  H(t),\cdot  \bigr]+ \sum_{\ell=1}^L \sum_{k,q}  \bigl( 
    L^\ell_{kq} \cdot L_{kq}^{\ell\dagger} -\frac{1}{2} \bigl\{L_{kq}^{\ell\dagger} L^\ell_{kq}   , \cdot\bigr\} \bigr) .
 \ee
Only considering the generator of the principle branch $ \mathcal{S}_0$, for $L=3$ and $L=4$, we obtain the maps shown in Fig.~\ref{fig:L3L4}(b) and (c), respectively.  Applying the Markovianity test to the unwound Floquet generator $\mathcal{S}_{\text{unw}}$ and $\mathcal{S}_0$ yields the phase diagrams in Fig.~\ref{fig:L3L4}(d) and (e), where we observe that the phase diagrams contain more valid Floquet Lindbladians $\mathcal{L}_\mathrm{F}$ (white region) \cite{Schnell2020,Schnell2021} or have significantly lower distances from Markovianity $\mu$ in the areas where no $\mathcal{L}_\mathrm{F}$ exists. Note again that %the alternative way of 
scanning through all the possible logarithm branches %(which is the alternative to undo the winding) 
would mean to scan through possibly $\approx 10^{14}$ candidates for $L=3$ and $\approx 10^{60}$ candidates for $L=4$, if we merely wanted to include the first two neighbouring branches $x_c \in \lbrace -1, 0, 1 \rbrace$. The unwound generator $\mathcal{S}_{\text{unw}}$ thus gives rise to a fast way of determining a particularly good candidate generator which can be tested for Markovianity, when  dealing with high-dimensional open quantum many-body systems. In %the SM \cite{SM} 
App.~\ref{sec:app-int} we provide further observations  for interacting systems  in terms of the existence of the Floquet Lindbladian.  Based on that, we motivate our specific choice of dissipation in this paper. In App.~\ref{sec:app-non} we further give an example to show that the unwound generator is a good candidate not only for Floquet systems but also in the general case of a time-dependent non-periodic map $\mathcal{V}(t)$. We show that for a convex combination 
\begin{equation}
    \mathcal{V}(t) = \lambda \mathcal{V}_1(t) + (1-\lambda) \mathcal{V}_2(t) ,
\end{equation}
where $\mathcal{V}_j(t) = e^{t\mathcal{L}_j}$ with Lindbladians $\mathcal{L}_j$ and $\lambda \in [0,1]$ (which previously has been studied in the context of definitions of Markovianity \cite{Chruscinski2010}) the unwound generator $\mathcal{S}_{\text{unw}}$ produces valid effective Lindbladians in areas where the generator of principle branch $\mathcal{S}_0$ does not.

%\emph{Summary.}---
\section{Summary}
We have demonstrated that the problem of deciding Markovianity \cite{Wolf_2008} (i.e.~deciding whether a given quantum channel was generated by a Lindbladian generator) can often be solved for cases with access to the full dynamical map. To this end,
we employ Floquet theory and perform a gauge transformation that minimizes the occurring micromotion. Our ideas also significantly simplify the calculation of effective Floquet-Lindbladians for complex many-body systems \cite{Schnell2020, Schnell2021, Volokitin2022} and can be applied in the context of Liouvillian learning \cite{Pastori2022} for trotterized open quantum systems.
Remarkably, already without dissipation, we have shown that the effective Floquet generator is generally not of von Neumann form.
This highlights the fact that, while for the Hamiltonian case any branch of the quasienergy spectrum yields a sensible description, for Liouvillians one has to be much more careful, since only some branches provide a valid physical description. 
\begin{acknowledgments}
%\emph{Acknowledgments.}---
This work was partially supported by the
Deutsche Forschungsgemeinschaft (DFG, German Research Foundation) via the Reasearch Unit FOR 5688
(Project No. 521530974). G.~D. acknowledges support from a fellowship of the German Academic Exchange Service (DAAD).
\end{acknowledgments}
%\nocite{*}

\appendix

\section{Phase diagram with and without dissipation}
\label{sec:app1}
\begin{figure*}
\includegraphics[width=0.65\linewidth]{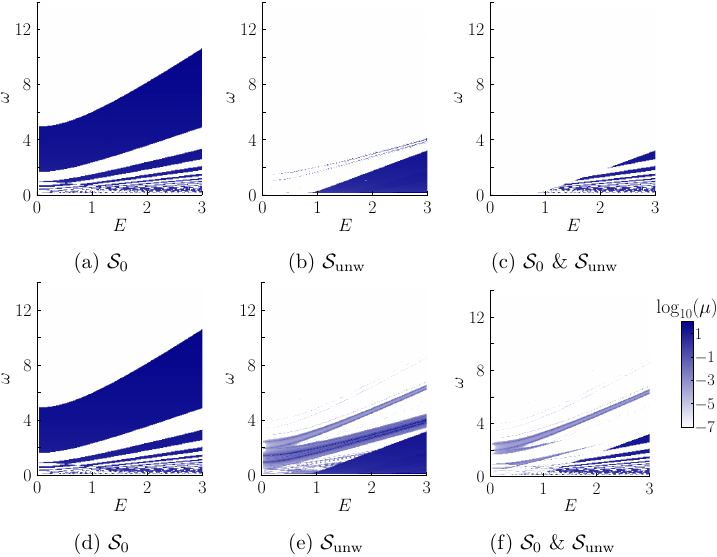} \hspace{0.0cm}
\caption{%\justifying
(a)-(f) Distance from Markovianity $\mu$ of the effective generator
of the one-cycle evolution superoperator as a function of driving
strength $E$ and frequency $\omega$, for a coupling strength $w=0.1$, a chain length of $L=2$ and (a)-(c) $\gamma=0$  and (d)-(f) $\gamma=0.01$. To lift the degeneracy, we set the energy splittings to $\Delta_1=1.01,\Delta_2=1.00,\Delta_3=0.98$.}
\label{fig:Comp}
\end{figure*}
In Fig.~\ref{fig:Comp} we show how the phase diagrams of $\mathcal{S}_{0}$ and $\mathcal{S}_{\text{unw}}$ complement each other in the coherent and dissipative case.

\begin{figure*}[]
\includegraphics[width=0.65\linewidth]{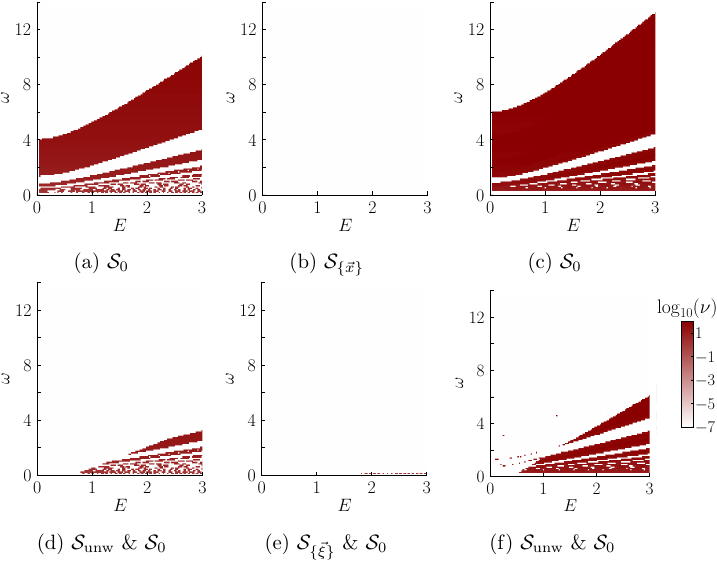} \hspace{0.0cm}
\caption{%\justifying
(a)-(f) Distance from von Neumann form $\nu$ of the effective generator
of the one-cycle evolution superoperator as a function of driving
strength $E$ and frequency $\omega$, for no dissipation,$\gamma=0$, coupling strength $w=0.1$, and chain lengths of (a),(b),(d),(e) $L=2$  and (c),(f) $L=3$. We set $\Delta_1=1.01,\Delta_2=1.00,\Delta_3=0.98$.}
\label{fig:vN}
\end{figure*}

\begin{figure*}
\includegraphics[width=0.8\linewidth]{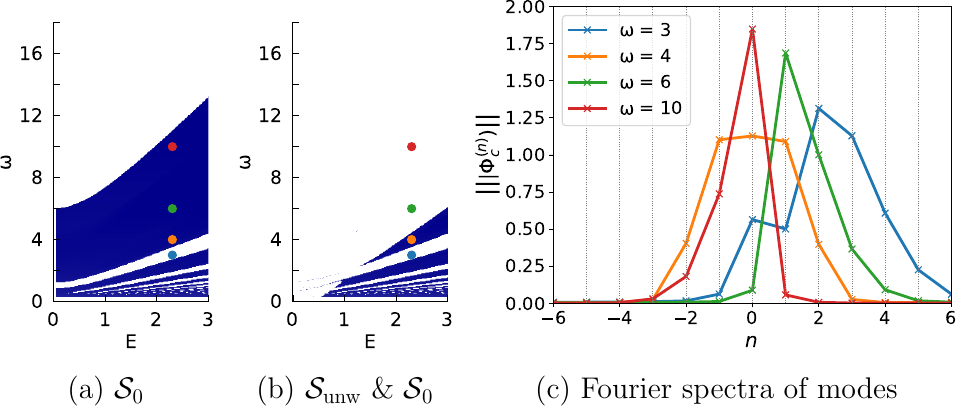} \hspace{0.0cm}
    \caption{\justifying
    (a) and (b) same as in Fig. 1(c) and (f) of the main text, respectively. (c) Fourier spectrum analysis of $|| \kett{\Phi_c^{(n)}} ||$ for a specific Floquet mode $c$, as a function of $n$. The different curves correspond to the points in (a) and (b). We observe that for $\omega=4$ (orange), where the procedure fails for $\mathcal{S}_\text{unw}$, $|| \kett{\Phi_c^{(-1)}} ||,|| \kett{\Phi_c^{(0)}} ||,|| \kett{\Phi_c^{(+1)}} ||$ are very close to each other such that here, all three $n$ should be considered. On the other hand, for the other frequencies $\omega$, for which the procedure works well, the amplitudes are peaked distinctly.}
    \label{fig:fouriermodes}
\end{figure*}

\section{Von-Neumann measure}
\label{sec:app-vN}

An arbitrary effective generator $\mathcal{S}_{\{\vec{x}\}}$ has von Neumann form if the left hand side of Eq.~\ref{eq:ccp} of the main text vanishes \cite{Schnell2021}
\be
    (\mathbf{1}_{d^2} - \Sigma )  \mathcal{S}_{\{\vec{x}\}}^\Gamma (\mathbf{1}_{d^2} - \Sigma ) \equiv  \mathcal{M}_{\{\vec{x}\}} = 0 .
\ee
Thus, to compute the minimal deviation from von Neumann form, we use the norm $||\mathcal{M}_{\{\vec{x}\}}||_1 = \Tr\sqrt{\mathcal{M}_{\{\vec{x}\}}^\dagger \mathcal{M}_{\{\vec{x}\}}}$ and introduce the measure
\be
\nu = \min_{\{\vec{x}\} \in  \mathbb{Z}^{N_c} }  \ \Bigl\{ ||\mathcal{M}_{\{\vec{x}\}}||_1 \Bigr\}  ,  \label{eq:vnmeasure}   
\ee
which we  interpret as ``distance from von Neumann form''. In Fig.~\ref{fig:vN} we can observe that we obtain the same zero-distance area in parameter space as in Fig.~\ref{fig:2spins} in the main text, whereas the precise values of the measures in the area of non-zero measure slightly deviate.
\begin{table*}
\includegraphics[width=1\linewidth]{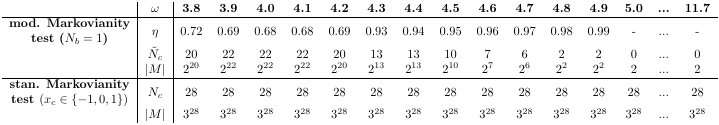} \hspace{0.0cm}
\caption{%\justifying
Comparison of the complexity of the modified and the standard Markovianity test for a spin chain of length $L=3$, driving strength of $E=2.5$ and no dissipation ($\gamma=0$). Note that for both methods we show the minimal cardinality $|M|$ which yields proper von Neumann generators for every point in the considered $\omega$-Intervall. }
\label{fig:tab}
\end{table*}

\section{Fourier spectrum analysis for the success of the spectral unwinding procedure}

\label{sec:app-Four}

The reason for the failure of the procedure at certain points in the phase diagram can be analyzed by investigating the Fourier spectra of the Floquet modes. An example is plotted in Fig.~\ref{fig:fouriermodes}, where in (a) and (b) we show the same phase diagrams as in Fig.~1(c) and (f) of the main text, respectively, and in (c) we analyze the Fourier spectra of $|| \kett{\Phi_c^{(n)}} ||$ for a specific Floquet mode $c$, as a function of $n$. The different curves correspond to the points in (a) and (b). We can see that for $\omega=3,6,10$, where the procedure gives good results for $\mathcal{S}_\text{unw}$ (compare (a) and (b)), the modes are distinctly peaked. However, for $\omega=4$ (orange), the procedure fails for $\mathcal{S}_\text{unw}$. Note that here $|| \kett{\Phi_c^{(-1)}} ||,|| \kett{\Phi_c^{(0)}} ||,|| \kett{\Phi_c^{(+1)}} ||$ are peaked similarly such that for this point in phase space, all three $n$ should be considered.

\section{Details on the complexity of the modified Markovianity test for $L=3$}

\label{sec:app-comp}

As mentioned in the main text, for $L\ge 3$ we cannot set $N_b$ and $\eta$ to any arbitrary values and expect to recover valid von Neumann generators for every single point in the phase diagram while also ensuring that the results can be obtained in reasonable computation times. Thus, for $L\ge 3$, for every point in the phase diagram one can pick an $N_b$ and $\eta$ and check if one obtains the correct branch combination. If yes (not), one can readjust the parameters by decreasing (increasing) $N_b$ and/or increasing (decreasing) $\eta$. As a consequence, this readjustment will decrease (increase) the complexity in terms of the computation time.  
In Fig.~\ref{fig:2spins}(e) of the main text (chain length $L=3$) we have set $N_b=1$ and varied $\eta$ for the vertical red line at $E=2.5$. The interval $\omega \in [3.8,11.7]$ that we consider, only contains points where the generator of the principle branch $\mathcal{S}_{0}$ does not yield a proper von Neumann generator. Starting at $\omega=3.8$ and using increments of $\Delta{\omega}=0.1$, for every point we have numerically determined the largest value of $\eta$ which yields a valid generator of von Neumann form. The results are shown in Tab.~\ref{fig:tab}. Note that for $\omega \in [5.0,11.7]$ the unwinded generator $\mathcal{S}_{\text{unf}}$ already constitutes the correct branch. 

\begin{figure}
\includegraphics[width=1\linewidth]{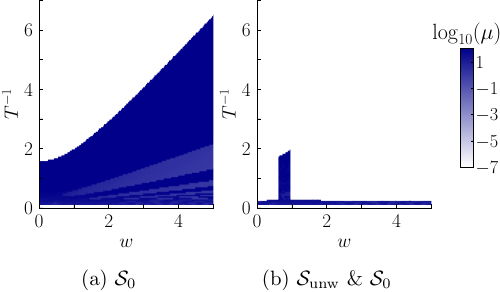} \hspace{0.0cm}
    \caption{\justifying
    Distance from Markovianity $\mu$ of the effective generator
of the evolution superoperator $\mathcal{V}(T)$ as a function of coupling strength $w$ and inverse observation time $T^{-1}$, for $\gamma = 0.01$, $\lambda=0.5$ a chain length of $L=3$. The energy splittings are set to $\Delta_1=2.45,\Delta_2=1.02,\Delta_3=1.65$.}
    \label{fig:Chrus}
\end{figure}

\begin{figure*}
\includegraphics[width=0.72\linewidth]{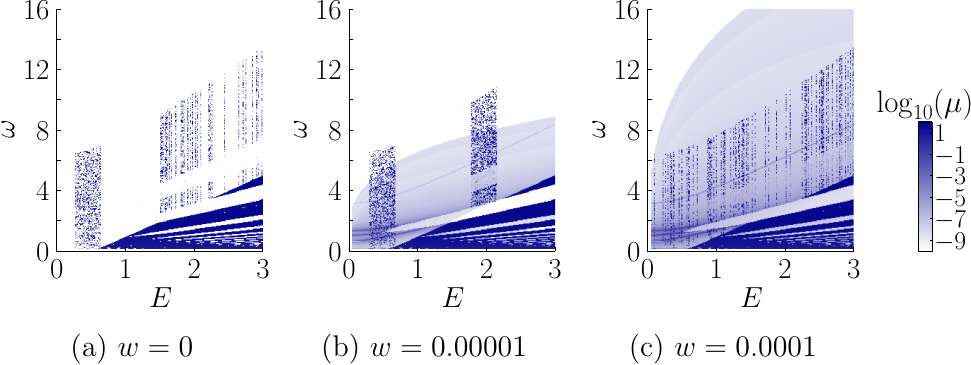} \hspace{0.0cm}
\caption{%\justifying
Distance from Markovianity $\mu$ of the effective generator
of the one-cycle evolution superoperator as a function of driving
strength $E$ and frequency $\omega$, for coupling strengths (a) $w=0$, (b) $w=0.00001$, (c) $w=0.0001$, a chain length of $L=3$, with dissipation from decay and a dissipation parameter of $\gamma=0.01$. Energy splittings are $\Delta_1=1.4,\Delta_2=1.1,\Delta_3=0.7$. For all plots we consider $\mathcal{S}_{0}$ and $\mathcal{S}_{\text{unw}}$.}
\label{fig:NonMarkBg}
\end{figure*}

In order to compare the complexity between the methods, we have quantified the complexity using the cardinality $|M|$ of the set $M$ of considered branches for the different methods. For the standard Markovianity test, considering solely $x_c \in \{ -1,0,1 \}$  (since this choice is enough to recover valid von Neumann generators everywhere), the size of $M$ is determined by the number of neighboring branches to the principal branch that are  included  and the number of complex conjugated eigenvalue pairs $N_c$. For every point $(E,\omega)$ the cardinality is given by $|M|=3^{N_c}$. For the modified Markovianity test, we first set $N_b=1$ (since in general this has given us the most efficient results) and varied $\eta$. The latter essentially sets a new number of `relevant' complex conjugated eigenvalue pairs $\Tilde{N}_c$ which represents the number of modes which are insufficiently peaked. Note that $0 \le \Tilde{N}_c  \le N_c$. The cardinality $|M|$ for this method is given by  $|M|=(N_b+1)^{\Tilde{N}_c} +1 $, where we add the $+1$ term for the generator of the principle branch $\mathcal{S}_{0}$, which we always check first. The results for $|M|$ for the different methods and points are shown in Tab.~\ref{fig:tab}. For the modified Markovianity test, the cardinality is on average $|M| \approx 10^5$ over the considered $\omega$-interval, which we plot in Fig.~\ref{fig:L3L4}(a) of the main text for $L=3$ (red).

\section{Phase diagrams for interacting systems and other forms of dissipation}
\label{sec:app-int}

Consider the Hamiltonian of the main text, i.e. a circularly driven spin chain of length $L$, 
\begin{align}
    & H(t) = \sum_{\ell=1}^L \frac{\Delta_\ell}{2}\sigma_z^\ell + \sum_{\ell=1}^{L-1}w\sigma_x^\ell \sigma_x^{\ell +1} + H_{\text{drive}}(t) \\ \label{eq:spinchain}
     & H_{\text{drive}}(t) =  \sum_{\ell=1}^L E \bigl[ \cos(\omega t) \sigma_x^\ell - \sin(\omega t) \sigma_y^\ell \bigr] ,
\end{align}
with level splittings $\Delta_\ell$ and coupling strengths $w$ between neighbouring spins. For a chain length of $L=3$ we add dissipation from decay to every spin according to the total Lindbladian
\be
    \mathcal{L}(t)[\cdot] = -i\bigl[H(t),\cdot \bigr]+ \sum_j \gamma \bigl(\sigma_-^j \cdot \sigma_+^j - \frac{1}{2} \{\sigma_+^j\sigma_-^j ,\cdot \} \bigr)  ,
    \label{eq:Lind-sigminus}
\ee
where $\gamma=0.01$.
For interacting systems we generally observe that even for very small coupling strengths $w$, we see a non-Markovian background arising and growing with increasing $w$, as we can see in Fig.~\ref{fig:NonMarkBg}.
Now recall that the most general form of a Lindbladian (i.e a generator of a quantum dynamical semigroup $\mathcal{V}(t)=e^{\mathcal{L}t}$) is given by~\cite{LME},\cite{Koss}
\be
    \mathcal{L}[\cdot] = -{i}\Bigl[  H,\cdot  \Bigr]+  \sum_{i,j=1}^{d^2-1} G_{ij} \biggl( F_{i} \cdot F_{j}^\dagger -\frac{1}{2} \Bigl\{F_{j}^\dagger F_{i}   , \cdot\Bigr\}    \biggr)  , \label{eq:LFApp}
\ee
where 
%we set $\hbar=k_\mathrm{B}=1$, $\{\cdot,\cdot \}$ denotes the anti-commutator, $H=H^\dagger$ the Hamiltonian, 
$G_{ij}$ a positive semidefinite Kossakowski matrix, and $F_{i}$ a basis of traceless operators. Note that for the model in Eq.~\eqref{eq:Lind-sigminus} the original Kossakowski matrix is very sparse, so that in the effective Lindbladian, many small but positive and negative eigenvalues of the effective Kossakowski matrix can be generated.
We therefore observe that the emergence of non-Markovian areas for even weakly interacting systems can be counteracted by allowing dissipation from several dissipation channels, i.e. high-rank Kossakowski matrices $G_{ij}$. Motivated by this observation, % to investigate open and interacting quantum systems, 
to avoid large non-Markovian regimes at finite interaction $w$, we chose the particular type of dissipation in the main 
\be
    \mathcal{L}(t)[\cdot] = -{i}\bigl[  H(t),\cdot  \bigr]+ \sum_{\ell=1}^L \sum_{k,q}  \bigl( 
    L^\ell_{kq} \cdot L_{kq}^{\ell\dagger} -\frac{1}{2} \bigl\{L_{kq}^{\ell\dagger} L^\ell_{kq}   , \cdot\bigr\} \bigr) ,
\ee
where for the specific form of the Lindblad jump operators $L_{kq}$ we refer to the main text.

\section{non-Floquet problem}
\label{sec:app-non}
To investigate and demonstrate the spectral unwinding procedure for a system that does not stem from a Floquet problem, we consider the Hamiltonian from the main text 
\be
     H = \sum_{\ell=1}^L \frac{\Delta_\ell}{2}\sigma_z^\ell + \sum_{\ell=1}^{L-1}w\sigma_x^\ell \sigma_x^{\ell +1} \label{eq:Hamilt}
\ee
but with $H_{\text{drive}}(t) = 0$. We now define two different Lindbladians 
\begin{align}
   & \mathcal{L}_1[\cdot] = -{i}\bigl[  H,\cdot  \bigr]+ \sum_{\ell=1}^L \gamma  \bigl( 
    \sigma_x^\ell \cdot \sigma_x^{\ell} -\frac{1}{2} \bigl\{\sigma_x^{\ell} \sigma_x^\ell  , \cdot\bigr\} \bigr) \\
   & \mathcal{L}_2[\cdot] = -{i}\bigl[  H,\cdot  \bigr]+ \sum_{\ell=1}^L \gamma  \bigl( 
    \sigma_y^\ell \cdot \sigma_y^{\ell} -\frac{1}{2} \bigl\{\sigma_y^{\ell} \sigma_y^\ell  , \cdot\bigr\} \bigr) .
\end{align}
We formally solve the respective master equations generated by the individual Lindbladians, 
\be
\frac{\partial}{\partial t}\rho = \mathcal{L}_j[\rho] ,
\ee
and extract the maps $\mathcal{V}_j(t) = e^{t\mathcal{L}_j}$. We now define the dynamical map (Ref.~\cite{Chruscinski2010})
\begin{equation}
    \mathcal{V}(t) = \lambda \mathcal{V}_1(t) + (1-\lambda) \mathcal{V}_2(t) ,
\end{equation} 

where $\lambda \in [0,1]$. We investigate its  dynamics up to a given observation time $T$.  To extract the effective generators and test them for Markovianitiy, we take the logarithm of $\mathcal{V}(T)$ and apply the Markovianity test outlined in the main text. In Fig.~\ref{fig:Chrus} we can see that the unwinded generator $\mathcal{S}_{\text{unw}}$ again produces valid effective Lindbladians in a large area where the generator of the principle branch $\mathcal{S}_{0}$ does not.

\bibliography{main}% Produces the bibliography via BibTeX.

\end{document}